\documentclass{article} 

\usepackage{spconf,epsfig,psfrag}
\usepackage{amsmath,amsfonts,amssymb,amsxtra,bm}
\usepackage{graphicx}
\usepackage{cite}

\newcommand{\eq}{\,=\,}
\newcommand{\be}{\begin{equation}}
\newcommand{\ee}{\end{equation}}
\newcommand{\ist}{\hspace*{.3mm}}
\newcommand{\rmv}{\hspace*{-.3mm}}

\allowdisplaybreaks

\title{Likelihood Consensus-Based Distributed Particle Filtering with\\[.7mm]Distributed Proposal Density Adaptation\vspace{.5mm}}

\name{Ondrej Hlinka$^1\rmv$, Franz Hlawatsch$^{1}\rmv$, and Petar M.\ Djuri\'c$^{\ist\ist 2}$
  \thanks{This work was supported by the Austrian Science Fund (FWF) under Award S10603, by the NSF under Award CCF-1018323, 
  and by the ONR under Award N00014-09-1-1154.}\vspace{-2.5mm}}
\address{\normalsize$^1$Institute of Telecommunications, Vienna University of Technology, Austria (ohlinka@nt.tuwien.ac.at)\\[-.5mm]
\normalsize $^2$Department of Electrical and Computer Engineering, Stony Brook University, NY, USA (djuric@ece.sunysb.edu)}

\ninept

\begin{document}

\maketitle

\renewcommand{\baselinestretch}{0.90}\small\normalsize

\begin{abstract}
We present a consensus-based distributed particle filter (PF) for 
wireless sensor networks. Each sensor runs a local PF to compute a global state estimate that takes into account the measurements of all sensors. 
The 
local PFs use the joint (all-sensors) likelihood function, which 
is calculated in a distributed way by a novel generalization of the likelihood consensus scheme.
A performance improvement (or a reduction of the required number of particles) is 
achieved by a novel distributed, con\-sensus-based method for adapting the 
proposal densities of the local PFs. 
The performance of the proposed distributed PF 
is demonstrated for a target tracking problem. 
\end{abstract}
%

\vspace{-3mm}

\section{Introduction}\label{sec:intro}

\vspace{-1mm}

We consider distributed sequential state estimation in a wireless sensor network.
For general nonlinear/non-Gaussian scenarios,
the particle filter (PF) is often
the estimation method of choice \cite{doucet2001sequential}. 
In this paper, extending our 
work in \cite{hlinka2010likelihoodcons,hlinka2011dgpf,hlinka2011journal}, 
we propose a distributed PF that uses a novel distributed scheme for proposal density adaptation.  
As in \cite{hlinka2010likelihoodcons,hlinka2011dgpf,hlinka2011journal}, each sensor runs a \emph{local} PF that computes a \emph{global} state estimate incorporating the measurements of \emph{all} sensors. 
The 
local PFs use the joint (all-sensors) likelihood function (JLF), which 
is computed
in a decentralized way by means of the \emph{likelihood consensus} (LC)
scheme. Here, we present a generalized form of the LC originally proposed in \cite{hlinka2010likelihoodcons},
which is suited to a general measurement model (i.e., it is not limited to additive Gaussian measurement noises \cite{hlinka2010likelihoodcons,hlinka2011dgpf}
or likelihoods from the exponential family \cite{hlinka2011journal}).

Our main contribution 
is a novel distributed, con\-sensus-based scheme for adapting the proposal densities (PDs) used by the local PFs.
Adapted PDs can yield a significant performance improvement 
or, alternatively, a significant reduction of the required number of particles \cite{simandl2007sampling}. 
In our adaptation scheme,
local PDs computed by the individual sensors are 
fused in a distributed way by means of consensus algorithms,
thereby providing to each local PF a global PD reflecting all measurements.
To make our 
scheme computationally feasible, we use Gaussian approximations for the local and global PDs. 
Our PD adaptation scheme differs from that proposed in \cite{van2003gaussian} in that it is distributed 
and it uses Gaussian
PD approximations.

Consensus-based distributed PFs with PD adaptation have been recently proposed in
\cite{farahmand2011set,mohammadi2011consensus}.  
The distributed PD adaptation scheme of \cite{farahmand2011set} employs min- and max-consensus 
to construct a set capturing most of the posterior probability mass. 
This set is used to calculate a distorted state-transition density, which serves as PD. 
Our distributed PD adaptation scheme has a lower complexity than that of \cite{farahmand2011set}. The 
communication requirements 
of our PD adaptation scheme are somewhat higher, but the overall communication requirements
of our distributed PF are still much lower than those of the distributed PF of \cite{farahmand2011set}, and 
simulation results demonstrate a better estimation performance of our distributed PF. In \cite{mohammadi2011consensus}, 
a distributed 
unscented PF is presented. This method employs a PD adaptation which, however,
is not distributed: the PD used at each sensor is only based on the local measurement. Again, simulation results demonstrate a better estimation performance 
of our distributed PF, which however comes at the cost of higher communication requirements.   

This paper is organized as follows. In Section \ref{sec:syst_mod}, we introduce the system model and review 
the principles of sequential Bayesian estimation. The LC-based distributed PF and the new generalized LC scheme are described in Section \ref{sec:DPF}. 
In Section \ref{sec:Distrib_Adapt}, we present the proposed distributed PD adaptation scheme. 
Finally, Section \ref{sec:sims} reports simulation results for a target tracking problem.

\vspace{-1mm}

\section{Sequential Bayesian State
estimation}\label{sec:syst_mod}

\vspace{-1mm}

We consider a random, time-varying 
state vector $\mathbf{x}_{n}=(x_{n,1}\cdots$\linebreak 
$x_{n,M})^\top\!$. The state 
evolves 
according to the state-transition model
\be
\mathbf{x}_{n} \!\eq\rmv \mathbf{g}_{n}(\mathbf{x}_{n-1},\mathbf{u}_{n}) \,, \quad\; n=1,2,\dots \,, 
\label{eq:stateModel_general}
\ee
where $\mathbf{u}_{n}$ is white driving noise with a known probability density function (pdf) $f(\mathbf{u}_{n})$.
At time $n$, 
$\mathbf{x}_{n}$ is sensed by a 
sensor network with $K$ sensors according to the measurement 
models
\be
\mathbf{z}_{n,k} \!\rmv\eq\rmv \mathbf{h}_{n,k}(\mathbf{x}_{n},\mathbf{v}_{n,k}) \,, \quad\; k=1,2,\dots, K \ist .
\label{eq:measModel_general}
\vspace{-.3mm}
\ee
Here, $\mathbf{z}_{n,k}$ of dimension $N_{n,k}$ is the measurement at time $n$ and at sensor $k$, 
and $\mathbf{v}_{n,k}$ is measurement noise with a known pdf $f(\mathbf{v}_{n,k})$. We assume that 
(i) $\mathbf{v}_{n,k}$ and $\mathbf{v}_{n'\rmv,k'}$ are independent unless $(n,k) \!=\! (n'\rmv,k')$;
(ii) the initial state $\mathbf{x}_{0}$ and the sequences
$\mathbf{u}_{n}$ and $\mathbf{v}_{n,k}$ are all independent; and (iii) sensor $k$ knows $\mathbf{g}_{n}(\cdot,\cdot)$ and $\mathbf{h}_{n,k}(\cdot,\cdot)$  
for all $n$, but it does not know $\mathbf{h}_{n,k'}(\cdot,\cdot)$ for $k'\!\!\not=\! k$. 

The state-transition and measurement models \eqref{eq:stateModel_general} and \eqref{eq:measModel_general} 
together with our statistical assumptions determine
the \emph{state-transition pdf} $\rmv f(\mathbf{x}_{n}|\mathbf{x}_{n-1})$, the \emph{local} likelihood function $f(\mathbf{z}_{n,k}|\mathbf{x}_{n})$,
and the JLF
$f(\mathbf{z}_{n}|\mathbf{x}_{n})$. Here, $\mathbf{z}_{n} \triangleq (\mathbf{z}_{n,1}^{\top} \rmv\cdots\ist \mathbf{z}_{n,K}^{\top})^{\top}\!$ 
denotes the vector containing all sensor measurements at time $n$.
Due to \eqref{eq:measModel_general} and the independence of all $\mathbf{v}_{n,k}$, the JLF is given 
\vspace{-.6mm}
by 
\be
f(\mathbf{z}_{n}|\mathbf{x}_n) \eq \prod_{k=1}^{K} f(\mathbf{z}_{n,k}|\mathbf{x}_{n}) \,. 
\label{eq:joint_likelihood}
\vspace{-.6mm}
\ee

Our goal is estimation of the state $\mathbf{x}_{n}$ from all sensor measurements from time $1$ to time $n$, 
$\mathbf{z}_{1:n} \triangleq ( \mathbf{z}_{1}^{\top} \!\cdots\ist \mathbf{z}_{n}^{\top} )^\top\!$.
To this end, we consider the minimum mean-square error (MMSE) 
\vspace{-.6mm}
estimator \cite{kay1998fundamentals}
\be
\label{eq:mmse}
\hat{\mathbf{x}}_{n}^{\text{MMSE}} \,\triangleq\, \text{E}\{\mathbf{x}_{n}|\mathbf{z}_{1:n}\}
\,= \int \rmv\mathbf{x}_{n} \ist f(\mathbf{x}_{n}|\mathbf{z}_{1:n}) \, d\mathbf{x}_{n} \,.
\vspace{-.6mm}
\ee
The posterior pdf $f(\mathbf{x}_{n}|\mathbf{z}_{1:n})$ in \eqref{eq:mmse}
can be calculated sequentially 
from the previous posterior 
$f(\mathbf{x}_{n-1}|\mathbf{z}_{1:n-1})$
and the JLF $f(\mathbf{z}_{n}|\mathbf{x}_{n})$ \cite{tanizaki1996nonlinear}. 
A computationally feasible approximation to this 
\pagebreak 
\emph{sequential MMSE state estimation}
is provided by the PF, which represents the posterior pdf $f(\mathbf{x}_n|\mathbf{z}_{1:n})$ by a set of weighted particles
 \cite{doucet2001sequential}. 

\vspace{-1mm}

\section{LC-based Distributed Particle Filter}\label{sec:DPF}

\vspace{-1mm}

The proposed LC-based distributed PF (LC-DPF) differs from 
our previous work in
\cite{hlinka2010likelihoodcons,hlinka2011dgpf,hlinka2011journal} 
by the generalized LC and the distributed PD adaptation (presented in Sections \ref{sec:principles-likelihood-cons} and \ref{sec:Distrib_Adapt}, respectively).
As in \cite{hlinka2010likelihoodcons,hlinka2011dgpf,hlinka2011journal}, each sensor 
tracks a particle representation 
of the global posterior $f(\mathbf{x}_n|\mathbf{z}_{1:n})$ using a \emph{local PF}.
At each time $n$, each local PF obtains a state estimate $\hat{\mathbf{x}}_{n,k}$ that is based on $\mathbf{z}_{1:n}$, i.e., \emph{all} sensor 
measurements up to time $n$.
This requires knowledge of the JLF $f(\mathbf{z}_{n}|\mathbf{x}_{n})$ 
as a function of 
$\mathbf{x}_{n}$.
An approximation of the JLF is provided to all sensors in a distributed way by means of the generalized LC. 
No communication between distant sensors or complex routing protocols are required.
Also, no particles, local state estimates, or measurements are communicated between the sensors. 
The proposed distributed PD adaptation scheme can yield a significant performance improvement or, alternatively, a significant reduction of the number of particles
(and, thus, of the computational complexity);
this comes at the cost of an increase in inter-sensor communications. 

\vspace{-1mm}


\subsection{Local PF Algorithm}\label{sec:LPF}

\vspace{-.5mm}

At a given time $n\!\ge\!1$, the local PF at sensor $k$ performs the following steps, which are identical for 
\vspace{-.5mm}
all $k$: 

\emph{Step 1}:\, A resampling \cite{doucet2001sequential} is applied to the $J$ particles 
\vspace{-.8mm}
$\big\{ \mathbf{x}_{n-1,k}^{(j)} \big\}_{j=1}^{J}$ with corresponding weights 
$\big\{ w_{n-1,k}^{(j)} \big\}_{j=1}^{J}$ (calculated at time $n\!-\!1$) 
that represent the previous global posterior $f(\mathbf{x}_{n-1}|\mathbf{z}_{1:n-1})$ at sensor $k$.  
This produces $J$ resampled particles 
\vspace{-1mm}
$\big\{ \bar{\mathbf{x}}_{n-1,k}^{(j)} \big\}_{j=1}^{J}$. 

\emph{Step 2}:\, Temporary particles $\big\{\mathbf{x}_{n,k}^{\prime (j)}\big\}_{j=1}^{J}$ are 
\vspace{-.8mm}
randomly drawn 
from $f(\mathbf{x}_{n}|\bar{\mathbf{x}}_{n-1,k}^{(j)}) \triangleq f(\mathbf{x}_{n}|\mathbf{x}_{n-1})\big|_{\mathbf{x}_{n-1} =\, \bar{\mathbf{x}}_{n-1,k}^{(j)}}\!$,
and a Gaussian approximation $\mathcal{N}(\mathbf{x}_n;{\bm{\mu}}'_{n,k},{\mathbf{C}}'_{n,k})$ 
of the ``predicted posterior''
\vspace{.3mm}
$f(\mathbf{x}_n|\mathbf{z}_{1:n-1})$ is calculated according 
\vspace{-1mm}
to
\begin{align}
&{\bm{\mu}}'_{n,k} \ist=\ist \frac{1}{J}\sum_{j=1}^{J}\mathbf{x}_{n,k}^{\prime (j)} \,, \quad\; 
{\mathbf{C}}'_{n,k} \ist=\ist \frac{1}{J}\sum_{j=1}^{J}\mathbf{x}_{n,k}^{\prime (j)}\mathbf{x}_{n,k}^{\prime (j) \top} 
  \!- {\bm{\mu}}'_{n,k}{\bm{\mu}}^{\prime \top}_{n,k} \,. \nonumber\\[-3mm]
&\label{eq_predpostgauss}\\[-7mm]
&\nonumber
\end{align}

\emph{Step 3} (jointly performed by all sensors, using communication with neighboring sensors):\, 
An adapted Gaussian PD $q(\mathbf{x}_n;\mathbf{z}_n) \triangleq \mathcal{N}(\mathbf{x}_n;{\bm{\mu}}_{n},{\mathbf{C}}_{n})$ 
involving
all sensor measurements is computed from 
the ${\bm{\mu}}'_{n,k'}$, ${\mathbf{C}}'_{n,k'}$, and $\mathbf{z}_{n,k'}$ ($k' \!= 1,\ldots,K$) 
by means of the distributed, consensus-based scheme described in Section \ref{sec:Distrib_Adapt}. 

\emph{Step 4}: \,
$J$ particles $\big\{\mathbf{x}_{n,k}^{(j)}\big\}_{j=1}^{J}$ are drawn from the 
\vspace{-.8mm}
PD $q(\mathbf{x}_n;\mathbf{z}_n)$. 

\vspace{.5mm}

\emph{Step 5} (jointly performed by all sensors, using communication with neighboring sensors):\, 
An approximation $\tilde{f}(\mathbf{z}_{n}|\mathbf{x}_{n})$  of the JLF $f(\mathbf{z}_{n}|\mathbf{x}_{n})$ 
is computed 
in a distributed way by means of the generalized LC described in Section \ref{sec:principles-likelihood-cons},
using 
the particles $\big\{\mathbf{x}_{n,k}^{(j)}\big\}_{j=1}^{J}$ drawn in Step 4. 

\emph{Step 6}:\, Weights associated with the 
particles $\mathbf{x}_{n,k}^{(j)}$ drawn in Step 4 are calculated according 
\vspace{-.5mm}
to
\[
w_{n,k}^{(j)} \eq \gamma \, \frac{\tilde{f}(\mathbf{z}_{n}|\mathbf{x}_{n,k}^{(j)}) \,f(\mathbf{x}_{n,k}^{(j)}|\mathbf{x}_{n-1,k}^{(j)})
}{q(\mathbf{x}_{n,k}^{(j)};\mathbf{z}_n)} \,, \quad\; j=1,\dots,J \,,
\vspace{-.5mm}
\] 
where $\gamma$ is chosen such that $\sum_{j=1}^J w_{n,k}^{(j)} = 1$.

\emph{Step 7}:\, From 
$\big\{\mathbf{x}_{n,k}^{(j)}, w_{n,k}^{(j)} \big\}_{j=1}^{J}$, an approximation of the global MMSE state estimate $\hat{\mathbf{x}}_{n}^{\text{MMSE}}$ in \eqref{eq:mmse} 
is computed according 
\vspace{-.5mm}
to 
\[
\hat{\mathbf{x}}_{n,k} \rmv\eq\rmv \sum_{j=1}^{J} w_{n,k}^{(j)} \ist \mathbf{x}_{n,k}^{(j)} \,.
\vspace{-.5mm}
\] 


\emph{Initialization}:\, The recursive procedure defined by 
\vspace{-.2mm}
Steps 1--7 is initialized at time $n\!=\!0$ by $J$ particles $\mathbf{x}_{0,k}^{(j)}$ 
randomly drawn 
\vspace{-.7mm}
from an appropriate prior pdf $f(\mathbf{x}_0)$, and by equal weights $w_{0,k}^{(j)} \equiv 1/J$.

\vspace{-1mm}


\subsection{Generalized Likelihood Consensus}\label{sec:principles-likelihood-cons}

\vspace{-.5mm}

We now present the generalized LC scheme that is used in Step 5 
to provide an approximate JLF to each sensor. In contrast to 
our previous work 
\cite{hlinka2010likelihoodcons,hlinka2011dgpf,hlinka2011journal}, this scheme is not limited to likelihoods with exponential form or to additive Gaussian measurement noises;
it is suitable for any type of likelihood and any measurement model \eqref{eq:measModel_general}.

To derive the generalized LC, we first take the logarithm of 
\eqref{eq:joint_likelihood}:
\be
\log f(\mathbf{z}_{n}|\mathbf{x}_n) \eq \sum_{k=1}^{K} \log f(\mathbf{z}_{n,k}|\mathbf{x}_{n}) \,. 
\label{eq:log_joint_likelihood}
\ee
Unfortunately, a consensus-based distributed calculation of 
\eqref{eq:log_joint_likelihood} 
is not possible in general because the terms of the sum depend on the unknown state $\mathbf{x}_n$. 
We therefore use the following approximate (finite-dimensional) basis expansions of the local log-likelihoods:
\be
\log f(\mathbf{z}_{n,k}|\mathbf{x}_{n}) \,\approx\, \sum_{r=1}^{R} \alpha_{n,k,r}(\mathbf{z}_{n,k}) \,{\varphi}_{n,r}(\mathbf{x}_n) \,.
\label{eq:logLocalLh_approx}
\ee
Here, $\alpha_{n,k,r}(\mathbf{z}_{n,k})$ are expansion
coefficients that contain all sensor-local information (including the sensor measurement $\mathbf{z}_{n,k}$) 
and $\varphi_{n,r}(\mathbf{x}_n)$ are fixed, sensor-independent basis functions that are assumed to be known to all sensors.
Substituting \eqref{eq:logLocalLh_approx} into \eqref{eq:log_joint_likelihood}, we obtain
\be
\log f(\mathbf{z}_{n}|\mathbf{x}_n) \,\approx\, \sum_{r=1}^{R} a_{n,r}(\mathbf{z}_n) \,{\varphi}_{n,r}(\mathbf{x}_n) \,, 
\label{eq:logJLF_approx2}
\vspace{-2mm}
\ee
with
\be
a_{n,r}(\mathbf{z}_n) \eq \sum_{k=1}^{K}  \alpha_{n,k,r}(\mathbf{z}_{n,k}) \,.
\label{eq:A}
\vspace{2mm}
\ee
The sum over all sensors in \eqref{eq:A} can be easily computed in a distributed way by means of a consensus 
algorithm \cite{OlfatiSaber07consensus} since the terms of the sum are real numbers (not functions of $\mathbf{x}_n$).

By exponentiating \eqref{eq:logJLF_approx2}, we finally obtain the following approximation of the JLF, denoted $\tilde{f}(\mathbf{z}_{n}|\mathbf{x}_n)$: 
\be
f(\mathbf{z}_{n}|\mathbf{x}_n) \ist\approx\ist \tilde{f}(\mathbf{z}_{n}|\mathbf{x}_n) 
   \ist=\ist \exp\! \Bigg( \sum_{r=1}^{R} a_{n,r}(\mathbf{z}_n) \,{\varphi}_{n,r}(\mathbf{x}_n) \!\Bigg) \ist.\,
\label{eq:JLF_approx}
\ee
Therefore, a sensor that knows the coefficients ${a}_{n,r}(\mathbf{z}_n)$ is able to evaluate the approximate JLF $\tilde{f}(\mathbf{z}_{n}|\mathbf{x}_n)$
for all values of $\mathbf{x}_n$. In fact, the vector of all coefficients, 
${\mathbf{a}}_n(\mathbf{z}_n) \triangleq \big(a_{n,1}(\mathbf{z}_n) \ist\cdots\ist a_{n,R}(\mathbf{z}_n) \big)^{\!\rmv\top}\!$, 
can be viewed as a \emph{sufficient statistic} \cite{kay1998fundamentals} that epitomizes the total measurement $\mathbf{z}_n$
within the limits of the approximation \eqref{eq:logLocalLh_approx}. 
The expressions \eqref{eq:A} and \eqref{eq:JLF_approx} allow a distributed, consensus-based calculation of $\tilde{f}(\mathbf{z}_n|\mathbf{x}_n)$ 
due to the following key facts. 
(i) The coefficients ${a}_{n,r}(\mathbf{z}_{n})$ do not depend on the state $\mathbf{x}_n$ but contain the information 
of all sensors (i.e., the expansion coefficients ${\alpha}_{n,k,r}(\mathbf{z}_{n,k})$ for all $k$). 
(ii) The state $\mathbf{x}_n$ enters into 
$\tilde{f}(\mathbf{z}_n|\mathbf{x}_n)$ only via the basis functions ${\varphi}_{n,r}(\cdot)$, which are sensor-independent and known to each sensor.
(iii) According to \eqref{eq:A}, the 
${a}_{n,r}(\mathbf{z}_{n})$ are sums in which each term contains only local information of a single sensor. 

At each time $n$, the expansion
coefficients ${\alpha}_{{r},n,k}(\mathbf{z}_{n,k})$ in \eqref{eq:logLocalLh_approx} are calculated locally at each sensor  
$k$ by means of \emph{least squares fitting} \cite{bjorck1996numerical}
based on the $J$ data points 
\vspace{-.4mm}
$\big\{ \mathbf{x}_{n,k}^{(j)} \ist, \log f(\mathbf{z}_{n,k}|{\mathbf{x}}_{n,k}^{(j)}) \big\}_{j=1}^{J}$.
Here, the use of the particles ${\mathbf{x}}_{n,k}^{(j)}$ drawn in Step 4 of the local PF algorithm
ensures a good approximation in those 
\pagebreak 
regions of the state-space where the approximate JLF 
$\tilde{f}(\mathbf{z}_{n}|\mathbf{x}_n)$ is evaluated in 
\vspace{-.4mm}
Step 6. 
(We assume that $f(\mathbf{z}_{n,k}|{\mathbf{x}}_{n,k}^{(j)}) \not= 0$.) 

The steps of the \emph{generalized LC scheme} performed at a given time $n$ can now be summarized as 
\vspace{.4mm}
follows: 

\emph{Step 1}:\, Sensor $k$ calculates the expansion
coefficients ${\alpha}_{{r},n,k}(\mathbf{z}_{n,k})$ in \eqref{eq:logLocalLh_approx}  
using least squares 
\vspace{.4mm}
fitting. 

\emph{Step 2}:\, The coefficients ${\alpha}_{{r},n,k}(\mathbf{z}_{n,k})$ of all sensors $k$ are added in a distributed way using a consensus algorithm. 
One instance of that
algorithm is employed for each $r \in \{ 1,\ldots,R\}$; all instances are executed in parallel.
After a sufficient number of consensus iterations, 
the $a_{{r},n}(\mathbf{z}_n)$ 
(see \eqref{eq:A}) for all ${r}$
are available at each 
\vspace{.4mm}
sensor. 

\emph{Step 3}:\, Using the $a_{{r},n}(\mathbf{z}_n)$, each sensor is able to 
\vspace{-.4mm}
evaluate the approximate 
JLF $\tilde{f}(\mathbf{z}_n|\mathbf{x}_n)$ 
for any value of $\mathbf{x}_n$ 
\vspace{.4mm}
according to \eqref{eq:JLF_approx}. 

The proposed LC-DPF 
(without the PD adaptation described in Section \ref{sec:Distrib_Adapt})
requires the transmission of $IR$
real numbers by each sensor
at each time $n$, where $I$ is the
number of consensus iterations performed by each consensus algorithm and $R$ 
(cf.\ \eqref{eq:logLocalLh_approx}) 
is the number of consensus algorithms executed in parallel.  
All transmissions are to neighboring sensors only, and their number does not depend on the measurement dimensions 
$N_{n,k}$. 
Thus,
the LC-DPF is particularly attractive in the case of high-dimensional measurements. 

\vspace{-1mm}

\section{Distributed proposal adaptation}\label{sec:Distrib_Adapt}

\vspace{-1.5mm}

We now present our 
distributed scheme for calculating the adapted
PD $q(\mathbf{x}_n;\mathbf{z}_n)$ (Step 3 of the local PF algorithm in Section \ref{sec:LPF}). This scheme can be summarized as follows. 
First, a ``pre-distorted'' local posterior 
is calculated at each sensor. The
local posteriors are then fused via 
a distributed fusion rule to obtain a global posterior, which is used as PD by each local 
PF.\footnote{Note 
that the global posterior used as PD is different from the global posterior that is obtained by the PF as described in 
Section \ref{sec:DPF}.} 
This PD takes into account the measurements of \emph{all} sensors, which is appropriate in view of the fact that 
the JLF is used in Step 6.
Our approach is inspired by the one from \cite{oreshkin2010async}, which however was proposed in a different context and uses a fusion rule different from ours.

We first note that the global posterior $f(\mathbf{x}_{n}|\mathbf{z}_{1:n})$ can be written (up to a normalization factor) as
\begin{align}
f(\mathbf{x}_{n}|\mathbf{z}_{1:n}) &\eq f(\mathbf{x}_{n} | \mathbf{z}_{1:n-1}, \mathbf{z}_{n}) \nonumber\\
&\,\propto\, f(\mathbf{z}_{n} | \mathbf{x}_{n}, \mathbf{z}_{1:n-1}) \, f(\mathbf{x}_{n}|\mathbf{z}_{1:n-1})\nonumber\\
&\eq f(\mathbf{z}_{n} | \mathbf{x}_{n}) \, f(\mathbf{x}_{n}|\mathbf{z}_{1:n-1})\nonumber\\
&\eq \Bigg[ \prod_{k=1}^{K} f(\mathbf{z}_{n,k}|\mathbf{x}_n) \Bigg] \, f(\mathbf{x}_{n}|\mathbf{z}_{1:n-1}) \,.
\label{eq:glob_pred_poster_0}
\end{align}
Let us suppose that each sensor $k$ calculates a (pre-distorted, nonnormalized) local pseudoposterior defined as
\be
\tilde{f}(\mathbf{x}_{n}|\mathbf{z}_{1:n-1},\mathbf{z}_{n,k})
\,\triangleq\, f(\mathbf{z}_{n,k}|\mathbf{x}_n) \, f^{1/K}(\mathbf{x}_{n}|\mathbf{z}_{1:n-1}) \,.
\label{eq:loc_pred_poster}
\ee
The product of all local pseudoposteriors equals the global posterior up to a factor:
\begin{align}
\hspace{-1.5mm}\prod_{k=1}^{K} \tilde{f}(\mathbf{x}_{n}|\mathbf{z}_{1:n-1},\mathbf{z}_{n,k})
&\ist= \Bigg[ \prod_{k=1}^{K} f(\mathbf{z}_{n,k}|\mathbf{x}_n) \Bigg] \, f(\mathbf{x}_{n}|\mathbf{z}_{1:n-1}) \label{eq:glob_pred_poster_1}\\[1mm]
\hspace{-1.5mm}&\ist\propto\ist f(\mathbf{x}_{n}|\mathbf{z}_{1:n}) \,,
\nonumber
\end{align}
where \eqref{eq:glob_pred_poster_0} has been used.
This posterior reflects all sensor measurements and could be employed as the global PD.
However, for
a simple distributed computation of \eqref{eq:glob_pred_poster_1}, we use Gaussian 
approximations of the local pseudoposteriors and the global posterior, i.e.,  
$\tilde{f}(\mathbf{x}_{n}|\mathbf{z}_{1:n-1},\mathbf{z}_{n,k}) \approx \mathcal{N}(\mathbf{x}_n;\tilde{\bm{\mu}}_{n,k},\tilde{\mathbf{C}}_{n,k})$ 
\pagebreak 
and 
$f(\mathbf{x}_{n}|\mathbf{z}_{1:n}) \approx q(\mathbf{x}_n;\mathbf{z}_n) \eq \mathcal{N}(\mathbf{x}_n;{\bm{\mu}}_{n},{\mathbf{C}}_{n})$. 
Then, using \eqref{eq:glob_pred_poster_1} and the rules for a product of Gaussian densities \cite{gales2006product}, 
we obtain the following expressions of the mean and covariance of the PD $q(\mathbf{x}_n;\mathbf{z}_n)$:
\be
\bm{\mu}_{n} =\ist \mathbf{C}_{n} \rmv \sum_{k=1}^{K} \tilde{\mathbf{C}}_{n,k}^{-1} \ist \tilde{\bm{\mu}}_{n,k} \,, \qquad
\mathbf{C}_{n} = \Bigg( \sum_{k=1}^{K} \tilde{\mathbf{C}}_{n,k}^{-1} \Bigg)^{\!\!-1} \rmv.
\label{eq:global_prop_mean_cov}
\ee
The sums over all sensors in these expressions can be easily calculated in a distributed way using consensus algorithms. 


To calculate the Gaussian approximation $\mathcal{N}(\mathbf{x}_n;\tilde{\bm{\mu}}_{n,k},\tilde{\mathbf{C}}_{n,k})$ of the local pseudoposterior 
$\tilde{f}(\mathbf{x}_{n}|\mathbf{z}_{1:n-1},\mathbf{z}_{n,k})$,
we note that \eqref{eq:loc_pred_poster} is the measurement 
update step of a 
Bayesian filter using the pre-distorted predicted posterior 
$f^{1/K}(\mathbf{x}_{n}|\mathbf{z}_{1:n-1})$ instead of the true predicted posterior $f(\mathbf{x}_{n}|\mathbf{z}_{1:n-1})$.   
Furthermore, each sensor calculated a Gaussian approximation 
of the predicted posterior, $f(\mathbf{x}_{n}|\mathbf{z}_{1:n-1}) \approx \mathcal{N}(\mathbf{x}_n;{\bm{\mu}}'_{n,k},{\mathbf{C}}'_{n,k})$
(see Step 2 in Section \ref{sec:LPF}); this entails the Gaussian approximation 
$f^{1/K}(\mathbf{x}_{n}|\mathbf{z}_{1:n-1}) \approx \mathcal{N}(\mathbf{x}_n;{\bm{\mu}}'_{n,k},K{\mathbf{C}}'_{n,k})$.
Since 
Gaussian models are thus used for both $\tilde{f}(\mathbf{x}_{n}|\mathbf{z}_{1:n-1},\mathbf{z}_{n,k})$ and $f^{1/K}(\mathbf{x}_{n}|\mathbf{z}_{1:n-1})$,
we propose to perform the measurement update in \eqref{eq:loc_pred_poster} by means of the update step of a \emph{Gaussian filter}
\cite{tanizaki1996nonlinear,van2004sigma,arasaratnam2009cubature,ito2000gaussian}. 
This is done locally at each sensor.

The operations of the proposed PD adaptation scheme performed at time $n$ 
can now be summarized as follows:

\emph{Step 1}:\, Each sensor $k$ computes the mean $\tilde{\bm{\mu}}_{n,k}$ and covariance $\tilde{\mathbf{C}}_{n,k}$ 
of the Gaussian approximation of the local pseudoposterior,
$\mathcal{N}(\mathbf{x}_n;\tilde{\bm{\mu}}_{n,k},\tilde{\mathbf{C}}_{n,k}) \approx \tilde{f}(\mathbf{x}_{n}|\mathbf{z}_{1:n-1},\mathbf{z}_{n,k})$.
This is done locally by performing a Gaussian filter update step with input mean $\bm{\mu}'_{n,k}$, input covariance $K\mathbf{C}'_{n,k}$, and 
measurement $\mathbf{z}_{n,k}$.
Here, $\bm{\mu}'_{n,k}$ and $\mathbf{C}'_{n,k}$ were obtained locally 
\vspace{.5mm}
according to \eqref{eq_predpostgauss}.

\emph{Step 2}:\, Consensus algorithms are used to calculate the sums over all sensors in \eqref{eq:global_prop_mean_cov}. 
This step requires communication with neighboring sensors. In total, $I \ist [ M+M(M \rmv+\rmv 1)/2 +\rmv1 ]$ real numbers are 
communicated by each sensor
at time $n$. Here, $I$ denotes the number of consensus iterations and $M$ denotes the dimension of the state $\mathbf{x}_n$.  
After convergence of the consensus algorithms, each sensor obtained the global PD 
\vspace{.5mm}
$q(\mathbf{x}_n;\mathbf{z}_n)=\mathcal{N}(\mathbf{x}_n;{\bm{\mu}}_{n},{\mathbf{C}}_{n})$. 







\vspace{-.5mm}

\section{Simulation Results}  \label{sec:sims} 

\vspace{-1mm}

We consider a target tracking application using acoustic amplitude sensors. 
The target is represented by the vector $\bm{\tau}_n=(x_{n} \,\ist y_{n} \,\ist \dot{x}_{n} \,\ist \dot{y}_{n})^{\top}$ containing the target's 2D position and 2D velocity 
in the $x$-$y$ plane. 
The vector $\bm{\tau}_n$ evolves with time $n$ according to $\bm{\tau}_{n} = \mathbf{G}\bm{\tau}_{n-1} + \mathbf{W}\mathbf{u}'_{n}$, $n=1,2,\dots\,$,
where the matrices $\mathbf{G}\in \mathbb{R}^{4\times 4}$ and $\mathbf{W}\in \mathbb{R}^{4\times 2}$ are chosen as in\cite{hlinka2011journal} 
and the $\mathbf{u}'_{n}$ are independent and identically distributed according to
$\mathbf{u}'_{n} \!\sim\! \mathcal{N}(\mathbf{0},\mathbf{C}_{u'})$ with $\mathbf{C}_{u'} \!=\rmv \mathrm{diag}(0.0033,0.0033)$.
The 
target motion model specified above is however assumed unknown to the simulated PFs.
Therefore, all simulated PFs use a random walk 
model $\mathbf{x}_n = \mathbf{x}_{n-1} + \mathbf{u}_n$, 
where the state $\mathbf{x}_n = (x_{n} \,\ist y_{n})^{\top}$ represents the position of the target and 
$\mathbf{u}_{n} \!\sim\! \mathcal{N}(\mathbf{0},\mathbf{C}_{u})$ with $\mathbf{C}_{u} \!=\rmv \mathrm{diag}(0.0528,0.0528)$ (cf. \eqref{eq:stateModel_general}).

\begin{figure*}[!]
\centering

\begin{minipage}[t]{0.31\textwidth}
\psfrag{CPF}[][][0.65]{\hspace{4mm}\raisebox{0mm}{CPF}}
\psfrag{LC-DPF}[][][0.65]{\hspace{7mm}\raisebox{0.5mm}{LC-DPF}}
\psfrag{DPF-1}[][][0.65]{\hspace{5mm}\raisebox{0.5mm}{DPF-1}}
\psfrag{DPF-2}[][][0.65]{\hspace{5mm}\raisebox{0mm}{DPF-2}}

\psfrag{n}[][][0.70]{\hspace{-3mm}\raisebox{-8mm}{ $n$}}
\psfrag{RMSE [m]}[][][0.70]{\hspace{0mm}\raisebox{10mm}{RMSE$_n$ [m]}}

\psfrag{0}[][][0.6]{\hspace{0mm}\raisebox{-4mm}{$0$}}
\psfrag{20}[][][0.6]{\hspace{0mm}\raisebox{-4mm}{$20$}}
\psfrag{40}[][][0.6]{\hspace{0mm}\raisebox{-4mm}{$40$}}
\psfrag{60}[][][0.6]{\hspace{0mm}\raisebox{-4mm}{$60$}}
\psfrag{80}[][][0.6]{\hspace{0mm}\raisebox{-4mm}{$80$}}
\psfrag{100}[][][0.6]{\hspace{0mm}\raisebox{-4mm}{$100$}}
\psfrag{120}[][][0.6]{\hspace{0mm}\raisebox{-4mm}{$120$}}
\psfrag{140}[][][0.6]{\hspace{0mm}\raisebox{-4mm}{$140$}}
\psfrag{160}[][][0.6]{\hspace{0mm}\raisebox{-4mm}{$160$}}
\psfrag{180}[][][0.6]{\hspace{0mm}\raisebox{-4mm}{$180$}}
\psfrag{200}[][][0.6]{\hspace{0mm}\raisebox{-4mm}{$200$}}

\psfrag{0.1}[][][0.6]{\hspace{-4mm}\raisebox{0mm}{$0.1$}}
\psfrag{0.2}[][][0.6]{\hspace{-4mm}\raisebox{0mm}{$0.2$}}
\psfrag{0.3}[][][0.6]{\hspace{-4mm}\raisebox{0mm}{$0.3$}}
\psfrag{0.4}[][][0.6]{\hspace{-4mm}\raisebox{0mm}{$0.4$}}
\psfrag{0.5}[][][0.6]{\hspace{-4mm}\raisebox{0mm}{$0.5$}}
\psfrag{0.6}[][][0.6]{\hspace{-4mm}\raisebox{0mm}{$0.6$}}

\centering
\hspace*{3mm}\includegraphics[height=3.5cm,width=.97\textwidth]{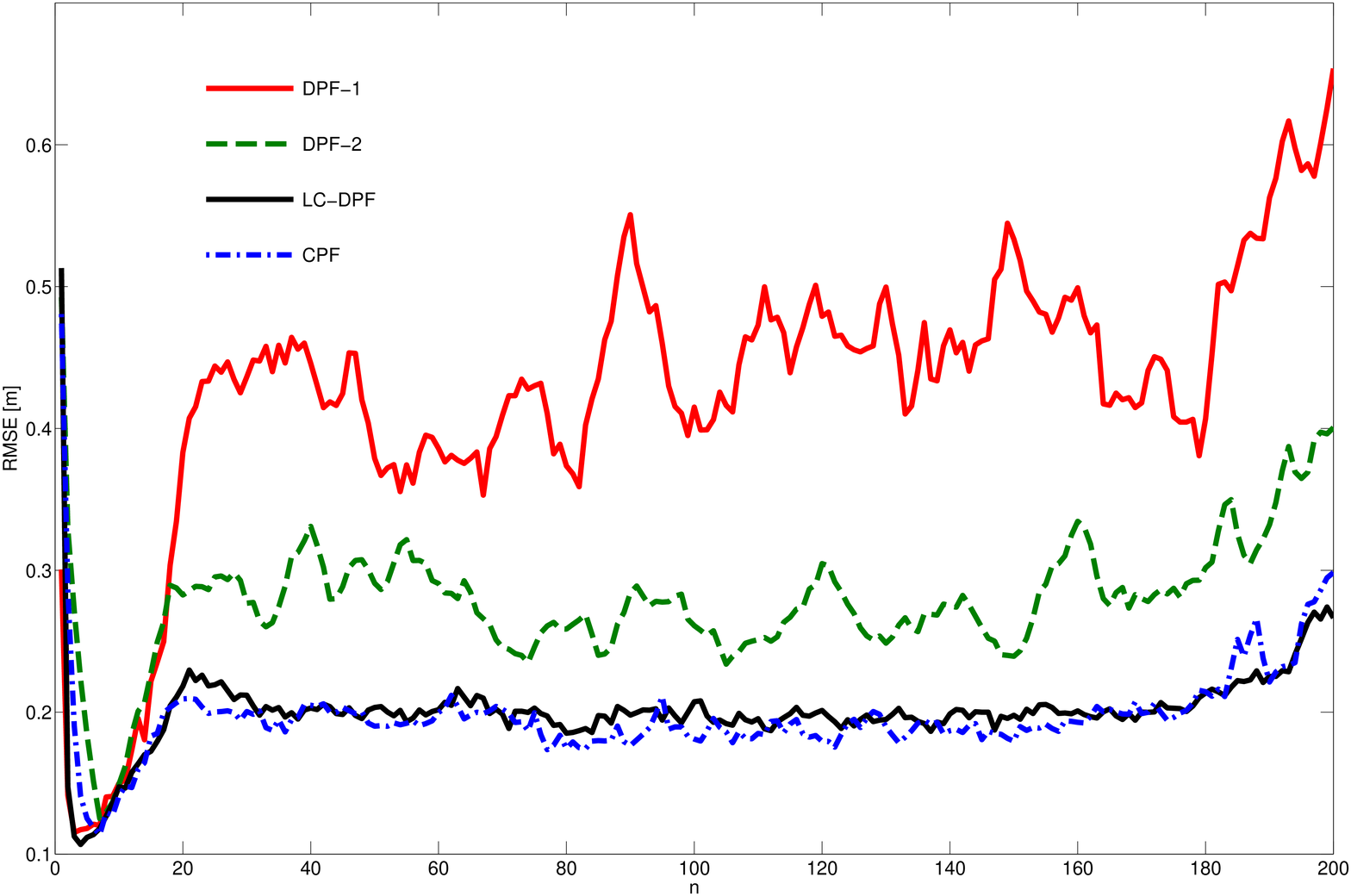}
\vspace*{-3.6mm}
\caption{RMSE$_n$ versus time $n$.}
\label{fig:RMSEvsTime}
\end{minipage}\hspace{5mm}
\begin{minipage}[t]{0.31\textwidth}
\psfrag{CPF}[][][0.65]{\hspace{4mm}\raisebox{0mm}{CPF}}
\psfrag{LC-DPF}[][][0.65]{\hspace{7mm}\raisebox{0.5mm}{LC-DPF}}
\psfrag{DPF-1}[][][0.65]{\hspace{5mm}\raisebox{0.5mm}{DPF-1}}
\psfrag{DPF-2}[][][0.65]{\hspace{5mm}\raisebox{0mm}{DPF-2}}

\psfrag{Order of polynomial approximation}[][][0.70]{\hspace{-3mm}\raisebox{-8mm}{$R_p$}}
\psfrag{ARMSE [m]}[][][0.70]{\hspace{0mm}\raisebox{9mm}{ARMSE [m]}}

\psfrag{1}[][][0.6]{\hspace{0mm}\raisebox{-4mm}{$1$}}
\psfrag{2}[][][0.6]{\hspace{0mm}\raisebox{-4mm}{$2$}}
\psfrag{3}[][][0.6]{\hspace{0mm}\raisebox{-4mm}{$3$}}
\psfrag{4}[][][0.6]{\hspace{0mm}\raisebox{-4mm}{$4$}}
\psfrag{5}[][][0.6]{\hspace{0mm}\raisebox{-4mm}{$5$}}
\psfrag{6}[][][0.6]{\hspace{0mm}\raisebox{-4mm}{$6$}}
\psfrag{7}[][][0.6]{\hspace{0mm}\raisebox{-4mm}{$7$}}
\psfrag{8}[][][0.6]{\hspace{0mm}\raisebox{-4mm}{$8$}}
\psfrag{9}[][][0.6]{\hspace{0mm}\raisebox{-4mm}{$9$}}
\psfrag{10}[][][0.6]{\hspace{0mm}\raisebox{-4mm}{$10$}}

\psfrag{0.15}[][][0.6]{\hspace{-4mm}\raisebox{0mm}{$0.15$}}
\psfrag{0.2}[][][0.6]{\hspace{-5mm}\raisebox{0mm}{$0.20$}}
\psfrag{0.25}[][][0.6]{\hspace{-4mm}\raisebox{0mm}{$0.25$}}
\psfrag{0.3}[][][0.6]{\hspace{-5mm}\raisebox{0mm}{$0.30$}}
\psfrag{0.35}[][][0.6]{\hspace{-4mm}\raisebox{0mm}{$0.35$}}
\psfrag{0.4}[][][0.6]{\hspace{-5mm}\raisebox{0mm}{$0.40$}}
\psfrag{0.45}[][][0.6]{\hspace{-4mm}\raisebox{0mm}{$0.45$}}
\psfrag{0.5}[][][0.6]{\hspace{-5mm}\raisebox{0mm}{$0.50$}}

\centering
\hspace*{1.3mm}\includegraphics[height=3.5cm,width=.97\textwidth]{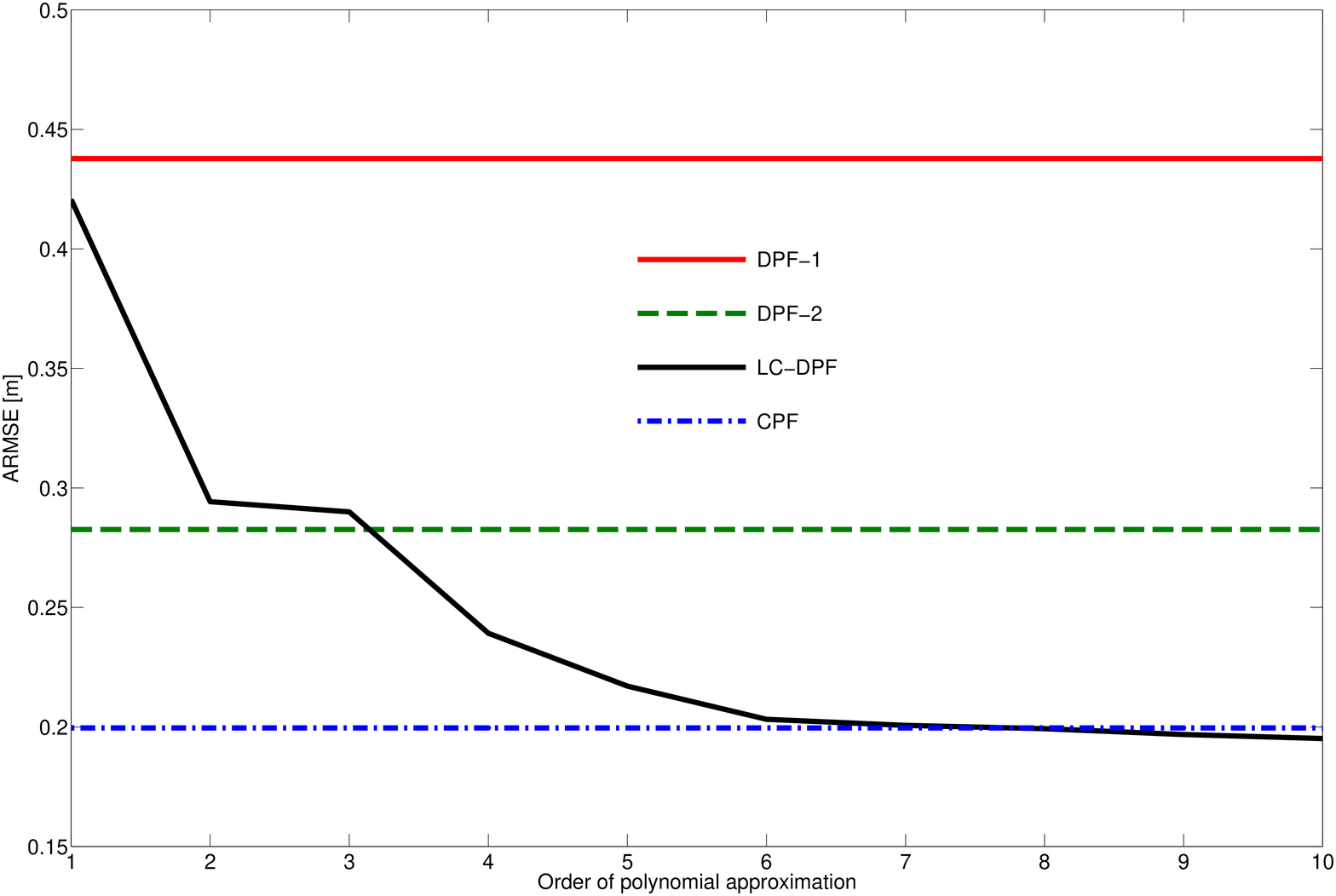}
\vspace*{-3.6mm}
\caption{ARMSE versus the degree of polynomial approximation $R_p$.} 
\label{fig:ARMSEvsPoly}
\end{minipage}\hspace{5mm}
\begin{minipage}[t]{0.31\textwidth}

\psfrag{LC-DPF-NA}[][][0.65]{\hspace{14mm}\raisebox{0mm}{LC-DPF-NA}}
\psfrag{LC-DPF}[][][0.65]{\hspace{10mm}\raisebox{0.5mm}{LC-DPF}}

\psfrag{Number of particles}[][][0.70]{\hspace{-3mm}\raisebox{-8mm}{$J$}}
\psfrag{ARMSE [m]}[][][0.70]{\hspace{0mm}\raisebox{10mm}{ARMSE [m]}}

\psfrag{0}[][][0.6]{\hspace{0mm}\raisebox{-4mm}{$ $}}
\psfrag{500}[][][0.6]{\hspace{0mm}\raisebox{-4mm}{$ $}}
\psfrag{1000}[][][0.6]{\hspace{0mm}\raisebox{-4mm}{$1000$}}
\psfrag{1500}[][][0.6]{\hspace{0mm}\raisebox{-4mm}{$ $}}
\psfrag{2000}[][][0.6]{\hspace{0mm}\raisebox{-4mm}{$2000$}}
\psfrag{2500}[][][0.6]{\hspace{0mm}\raisebox{-4mm}{$ $}}
\psfrag{3000}[][][0.6]{\hspace{0mm}\raisebox{-4mm}{$3000$}}
\psfrag{3500}[][][0.6]{\hspace{0mm}\raisebox{-4mm}{$ $}}
\psfrag{4000}[][][0.6]{\hspace{0mm}\raisebox{-4mm}{$4000$}}
\psfrag{4500}[][][0.6]{\hspace{0mm}\raisebox{-4mm}{$ $}}
\psfrag{5000}[][][0.6]{\hspace{0mm}\raisebox{-4mm}{$5000$}}

\psfrag{0.1}[][][0.6]{\hspace{-4.5mm}\raisebox{0mm}{$0.1$}}
\psfrag{0.2}[][][0.6]{\hspace{-4.5mm}\raisebox{0mm}{$0.2$}}
\psfrag{0.3}[][][0.6]{\hspace{-4.5mm}\raisebox{0mm}{$0.3$}}
\psfrag{0.4}[][][0.6]{\hspace{-4.5mm}\raisebox{0mm}{$0.4$}}
\psfrag{0.5}[][][0.6]{\hspace{-4.5mm}\raisebox{0mm}{$0.5$}}
\psfrag{0.6}[][][0.6]{\hspace{-4.5mm}\raisebox{0mm}{$0.6$}}
\psfrag{0.7}[][][0.6]{\hspace{-4.5mm}\raisebox{0mm}{$0.7$}}
\psfrag{0.8}[][][0.6]{\hspace{-4.5mm}\raisebox{0mm}{$0.8$}}
\psfrag{0.9}[][][0.6]{\hspace{-4.5mm}\raisebox{0mm}{$0.9$}}
\psfrag{1}[][][0.6]{\hspace{-4.5mm}\raisebox{0mm}{$1.0$}}
\psfrag{1.1}[][][0.6]{\hspace{-4.5mm}\raisebox{0mm}{$1.1$}}

\centering
\includegraphics[height=3.5cm,width=.97\textwidth]{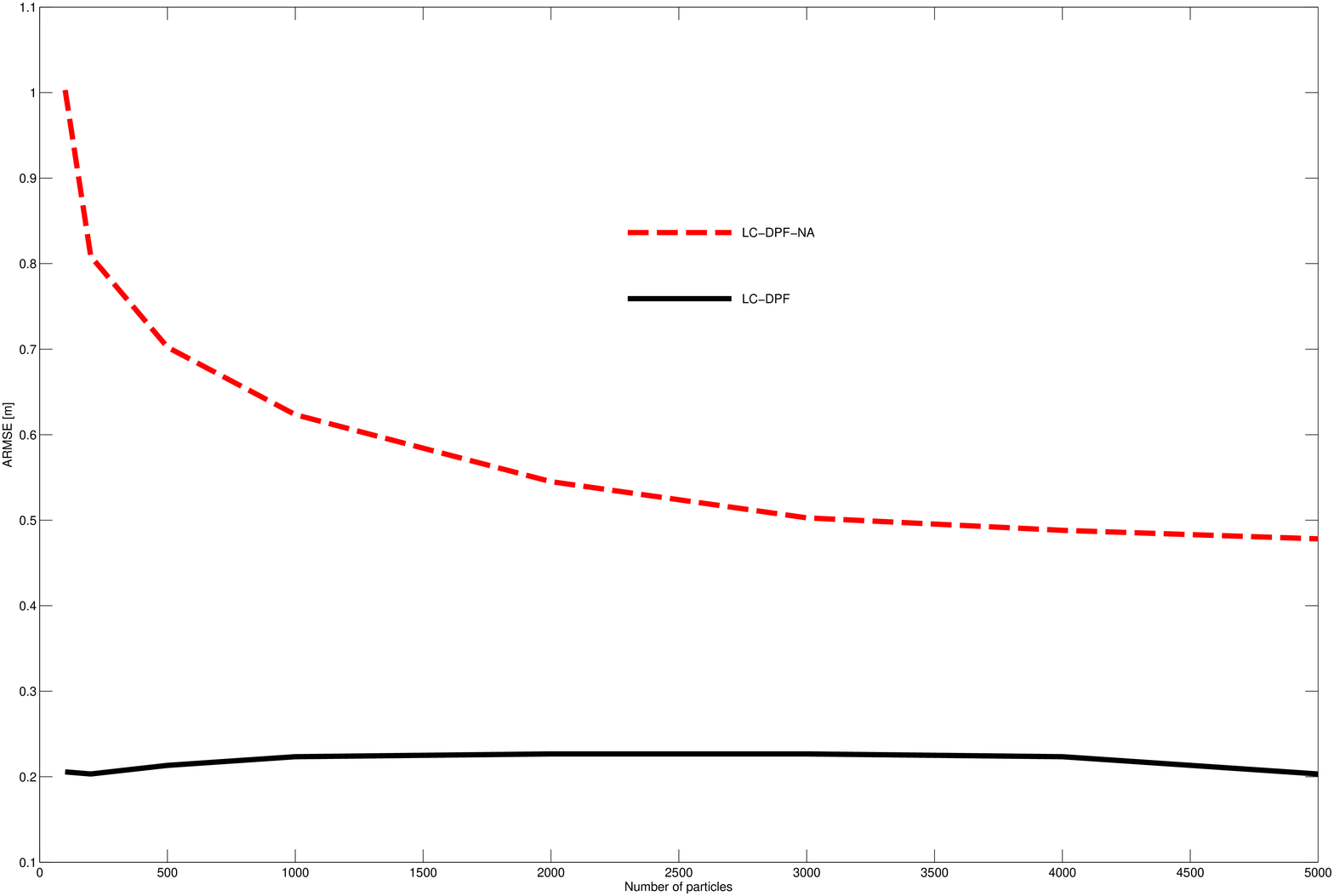}
\vspace*{0mm}
\caption{ARMSE versus the number of particles $J$.} 
\label{fig:ARMSEvsPart}
\end{minipage}
\vspace{-1mm}
\end{figure*}

The target emits a sound of constant amplitude $A \!=\! 10$, which is sensed by acoustic amplitude sensors. 
The (scalar) measurement $z_{n,k}\rmv$ of sensor $k$ is given by 
\vspace{-.5mm}
(cf.\ \eqref{eq:measModel_general})
\[
z_{n,k} \rmv\eq 
\frac{A}{\|\mathbf{x}_n \!-\bm{\xi}_{n,k} \|^{2}} \ist+\ist v_{n,k} \,,
\vspace{-.1mm}
\]
where 
$\bm{\xi}_{n,k}\rmv$ is
the position of sensor $k$ at time $n$ and $v_{n,k} \rmv\sim\mathcal{N}(0,\sigma_{v}^{2})$ with $\sigma_{v}^{2}=0.00005$.
This value of $\sigma_{v}^{2}$ yields a peaky likelihood, which highlights the performance gains of PD adaptation. 
The network consists
of $K \!=\! 25$ sensors that are deployed on a jittered grid 
within a rectangular region of size 
$40\ist\text{m} \times 40\ist\text{m}$. Each sensor communicates with other sensors within a range of $18\ist$m.

For LC, unless stated otherwise, we approximate 
$\log f(\mathbf{z}_{n,k}|\mathbf{x}_{n})$ 
by a multivariate polynomial of degree $R_p=6$; this leads to 
a basis expansion \eqref{eq:logLocalLh_approx} of degree $R \rmv=\! \binom{R_p+M}{R_p} \!=\rmv 28$.
The sums 
in \eqref{eq:A} and \eqref{eq:global_prop_mean_cov} are computed by $I \!=\! 15$ iterations of an average consensus algorithm with Metropolis weights 
\cite{xiao2005scheme}. For 
PD adaptation (Step 1 in Section \ref{sec:Distrib_Adapt}), the update step of an unscented Kalman filter \cite{van2004sigma} is used. 

We compare the proposed LC-DPF with the 
distributed PFs presented in \cite{farahmand2011set} and \cite{mohammadi2011consensus} 
(referred to as DPF-1 and DPF-2, respectively) 
and with a centralized PF (CPF) that processes all sensor measurements 
at a fusion center. The CPF 
uses an adapted PD that is computed 
using a (centralized) unscented Kalman filter. 
The number of particles at each sensor of the distributed PFs and at the fusion center of the CPF is $J \!=\! 200$ unless stated otherwise. 
As a performance measure, we use the 
root-mean-square error of the state estimates $\hat{\mathbf{x}}_{n,k}$, denoted $\text{RMSE}_n$, which is computed as the square root of the 
average of the squared estimation error over all sensors and over $1000$ simulation runs. 
We also compute the \emph{average} RMSE (ARMSE) 
by averaging $\text{RMSE}_n^2$ over all $200$ simulated time instants $n$ and taking the square root of the result.

Fig.\ \ref{fig:RMSEvsTime} shows the temporal 
evolution of $\text{RMSE}_n$. It can be seen that the performance of LC-DPF is almost as good as that of CPF and 
better than that of DPF-1 and DPF-2. The communication requirements of LC-DPF are lower than those of DPF-1 but higher than those of DPF-2:
the total counts of real numbers transmitted by LC-DPF, DPF-1, and DPF-2 during one time step in the entire network (all sensors) are 12375, 76875, and 1875, respectively. 

Fig.\ \ref{fig:ARMSEvsPoly} shows the ARMSE versus the degree $R_p$ of the polynomial used to approximate the local log-likelihood functions. 
As expected, the ARMSE of LC-DPF decreases with growing $R_p$ and approaches that of CPF. 
Note, however, that the communication requirements increase with growing $R_p$. 

Finally, Fig.\ \ref{fig:ARMSEvsPart} shows the dependence of the ARMSE on the number $J$ of particles for LC-DPF and an LC-DPF 
without proposal adaptation (abbreviated as LC-DPF-NA) \cite{hlinka2011journal}.  
As we can see, the ARMSE of LC-DPF is significantly lower than that of LC-DPF-NA, even if $J$ is large.
This demonstrates the performance improvement achieved by our distributed PD adaptation scheme.

\vspace{-1.5mm}

\section{Conclusion}

\vspace{-1.8mm}

We presented a consensus-based distributed particle filter (PF) for wireless sensor networks.
The state estimates computed by the local PFs at the various sensors reflect the past and present measurements of all sensors.
This is enabled by a generalized likelihood consensus scheme, which performs a distributed approximate calculation of 
the joint 
likelihood function for general measurement models. Our main contribution was
a 
distributed method for adapting the proposal density (PD) used by the local PFs. This method is based on a Gaussian model for the PD,
whose mean and covariance are computed in a distributed way
by consensus algorithms and by the update step of a Gaussian filter.  
Simulation results 
\pagebreak 
demonstrated the good performance of the proposed distributed PF
and the large performance gains achieved by the proposed PD adaptation method.


\renewcommand{\baselinestretch}{0.84}\normalsize\footnotesize

\bibliographystyle{ieeetr}
\bibliography{references}

\end{document}